\documentclass[10pt]{iopart}
\usepackage{iopams}
\usepackage{setstack}
\usepackage{amsthm}
\usepackage{graphicx}

\renewcommand{\vec}{\boldsymbol}
\renewcommand{\leq}{\leqslant}
\renewcommand{\geq}{\geqslant}
\renewcommand{\mathindent}{0.65in}

\eqnobysec

\newtheorem{conjecture}{Conjecture}[section]

\begin{document}

\title{Asymptotic dynamics of the exceptional Bianchi cosmologies}
\author{C G Hewitt\dag\ddag, J T Horwood\dag~and J Wainwright\dag}
\address{\dag Department of Applied Mathematics, University of Waterloo,
  Waterloo, Ontario, Canada N2L 3G1}
\address{\ddag St.~Jerome's University, Waterloo, Ontario, Canada N2L~3G3}

\begin{abstract}
In this paper we give, for the first time, a qualitative description of the
asymptotic dynamics of a class of non-tilted spatially homogeneous (SH)
cosmologies, the so-called exceptional Bianchi cosmologies, which are of
Bianchi type VI$_{-1/9}$. This class is of interest for two reasons. Firstly,
it is generic within the class of non-tilted SH cosmologies, being of the same
generality as the models of Bianchi types VIII and IX. Secondly, it is the
SH limit of a generic class of spatially inhomogeneous $G_{2}$ cosmologies.

Using the orthonormal frame formalism and Hubble-normalized variables, we show
that the exceptional Bianchi cosmologies differ from the non-exceptional
Bianchi cosmologies of type VI$_{h}$ in two significant ways. Firstly, the
models exhibit an oscillatory approach to the initial singularity and hence
are not asymptotically self-similar. Secondly, at late times, although the
models are asymptotically self-similar, the future attractor for the
vacuum-dominated models is the so-called Robinson-Trautman SH model instead of
the vacuum SH plane wave models.
\end{abstract}

\pacs{0420H, 0440N, 9880H}

%%%%%%%%%%%%%%%%%%%%%%%%%%%%%%%%%%%%%%%%%%%%%%%%%%%%%%%%%%%%%%%%%%%%%%%%%%%

\section{Introduction} \label{sec_intro}

In this paper we give a qualitative description of the asymptotic dynamics of
a class of spatially homogeneous (SH) cosmological models which to date have
received little attention in the literature. The matter content is a perfect
fluid with equation of state $p = (\gamma-1)\mu$, where $\gamma$ is a
constant. The models are \textit{non-tilted}, i.e.\ the fluid 4-velocity is
orthogonal to the orbits of the 3-parameter group $G_{3}$ of isometries, that
is of Bianchi type VI$_{h}$, with the value of the group parameter
$h = -\frac{1}{9}$. This exceptional value permits the shear tensor to have an
additional independent component (see Ellis and MacCallum~1969, p~123,
case~Bbii) with the result that this class of models depends on five arbitrary
parameters instead of four, making it of maximum generality, i.e.\ of the same
generality as the non-tilted Bianchi VIII and Bianchi IX models. We shall
refer to this class of SH cosmologies as the \textit{exceptional Bianchi
cosmologies}.

The longer term goal in studying the dynamics of SH cosmologies is to gain
insight into the dynamics of spatially inhomogeneous cosmologies. Indeed our
interest in the exceptional Bianchi cosmologies stems from the fact that
they are closely related to the simplest class of spatially inhomogeneous
cosmologies, the so-called $G_{2}$ cosmologies, which admit a 2-parameter
Abelian isometry group $G_{2}$. The relation is due to the fact that the
present class\footnote[1]{Models of all Bianchi types except for types VIII
and IX have this property.} admits an Abelian $G_{2}$ of isometries as a
subgroup of the full isometry group. Most research on $G_{2}$ cosmologies has
dealt with two special cases, either diagonal $G_{2}$ cosmologies, in which
case the line element is diagonal, or more generally, orthogonally transitive
$G_{2}$ cosmologies, in which case the line element has a \mbox{$2 \times 2$}
block form (see van~Elst \etal 2002, where other references can be found).
The exceptional Bianchi cosmologies are unique among the non-tilted SH
cosmologies in that \textit{the $G_{2}$ subgroup does not act orthogonally
transitively}, and, as a result, this class of models is the SH
limit\footnote[1]{There are also classes of \textit{tilted} SH models that are
SH limits of the most general $G_{2}$ cosmologies (see Wainwright and Ellis
1997, table~12.4, p~268).} of the most general $G_{2}$ cosmologies. One thus
expects that the asymptotic dynamics of the present class of SH cosmologies
will provide insight into the dynamics of general $G_{2}$ cosmologies.

In this paper we show that the presence of the additional shear degree of
freedom has a significant effect on the asymptotic dynamics of the
exceptional Bianchi cosmologies, both in the singular regime and in the
late-time regime. Firstly, this shear degree of freedom destabilizes the
vacuum Kasner solutions, leading to an oscillatory singularity. Secondly, it
destabilizes the SH vacuum plane wave solutions, with the result that they no
longer play a dominant role as regards the late-time dynamics.

The plan of this paper is as follows. In \sref{sec_eveqn} we present the
evolution equations for the exceptional Bianchi cosmologies, using the
orthonormal frame formalism, and discuss some properties of the
Hubble-normalized state space. Details of the derivation are provided in
appendix~A. In \sref{sec_eqp} we give the equilibrium points of the evolution
equations, and describe their local stability, indicating the various
bifurcations that are determined by the equation of state parameter $\gamma$.
Sections~\ref{sec_late} and \ref{sec_past} contain the main results of the
paper, a description of the asymptotic dynamics of the models at late times
and near the initial singularity, in terms of the future attractor and the past
attractor of the evolution equations. Finally, in \sref{sec_discussion} we
conclude by pointing out some analogies between the exceptional Bianchi
cosmologies and other classes of SH cosmologies with an
oscillatory singularity. We also comment on the implications of our results
for $G_{2}$ cosmologies.

It is assumed that the reader is familiar both with the orthonormal frame
formalism of Ellis and MacCallum (see MacCallum~1973), and with the use of
dynamical systems techniques in cosmology, for which we refer the
reader to the book of Wainwright and Ellis~(1997)\footnote[2]{Henceforth, we
will refer to this work as WE.}. We use geometrized units with
$8 \pi G = c = 1$ and the sign conventions of MacCallum~(1973).

%%%%%%%%%%%%%%%%%%%%%%%%%%%%%%%%%%%%%%%%%%%%%%%%%%%%%%%%%%%%%%%%%%%%%%%%%%%

\section{Evolution equations} \label{sec_eveqn}

In this section we give the evolution equations for the exceptional Bianchi
cosmologies. The equations are derived in appendix~A using the orthonormal
frame formalism and Hubble-normalized variables. The models are described by a
dimensionless state vector
\begin{equation}
  \vec{x} = ( \Sigma_{+}, \Sigma_{-}, \Sigma_{2}, \Sigma_{\times}, N_{-},
    A ), \label{eq_eveqn_statevector}
\end{equation}
subject to a constraint given by
\begin{equation}
  g(\vec{x}) = (\Sigma_{+} + \sqrt{3}\, \Sigma_{-})A - \Sigma_{\times}
    N_{-} = 0 . \label{eq_eveqn_cons}
\end{equation}
The variables $\Sigma_{+}$, $\Sigma_{-}$, $\Sigma_{2}$ and $\Sigma_{\times}$
describe the shear of the fluid congruence, while $N_{-}$ and $A$ describe the
spatial curvature. These variables are dimensionless, having been normalized
with the Hubble scalar\footnote[1]{On account of \eref{eq_eveqn_H},
$H$ is related to the rate of volume expansion $\Theta$ of the fluid
congruence according to $H = \frac{1}{3} \Theta$.} $H$, which is related to
the overall length scale $\ell$ by
\begin{equation}
  H = \frac{ \dot{\ell} }{ \ell } . \label{eq_eveqn_H}
\end{equation}
The overdot denotes differentiation with respect to clock time along
the fluid congruence. The state variables depend on a dimensionless
time variable $\tau$ that is related to the length scale $\ell$ by
\begin{equation}
  \ell = \ell_{0} \rme^{\tau} , \label{eq_eveqn_ell}
\end{equation}
where $\ell_{0}$ is a constant. The dimensionless time $\tau$ is related
to the clock time $t$ by
\begin{equation}
  \frac{\rmd t}{\rmd \tau} = \frac{1}{H}, \label{eq_eveqn_t_tau}
\end{equation}
as follows from equations \eref{eq_eveqn_H} and \eref{eq_eveqn_ell}.
In formulating the evolution equations we require the deceleration parameter
$q$, defined by
\begin{equation}
  q = -\frac{\ell \ddot{\ell}}{\dot{\ell}^{2}} , \label{eq_eveqn_q_defn}
\end{equation}
and the density parameter $\Omega$, defined by
\begin{equation}
  \Omega = \frac{\mu}{3 H^{2}}. \label{eq_eveqn_Omega_defn}
\end{equation}

The evolution equations for the components of $\vec{x}$ in
\eref{eq_eveqn_statevector} take the following form
\begin{eqnarray}
  \eqalign{
  \Sigma'_{+} &= (q-2)\Sigma_{+} + 3 \Sigma_{2}^{2} - 2 N_{-}^{2} - 6 A^{2}, \\
  \Sigma'_{-} &= (q-2)\Sigma_{-} - \sqrt{3}\, \Sigma_{2}^{2} + 2\sqrt{3}\,
    \Sigma_{\times}^{2} - 2\sqrt{3}\, N_{-}^{2} + 2\sqrt{3}\, A^{2}, \\
  \Sigma'_{2} &= (q-3\Sigma_{+}+\sqrt{3}\,\Sigma_{-}-2) \Sigma_{2} , \\
  \Sigma'_{\times} &= (q-2\sqrt{3}\,\Sigma_{-}-2)\Sigma_{\times} -
    8 N_{-} A , \\
  N'_{-} &= (q+2\Sigma_{+}+2\sqrt{3}\,\Sigma_{-})N_{-} + 6 \Sigma_{\times} A \\
  A' &= (q+2\Sigma_{+}) A , } \label{eq_eveqn_eveqn}
\end{eqnarray}
where
\begin{eqnarray}
  q &= 2 \Sigma^{2} + \case{1}{2}(3\gamma-2)\Omega , \label{eq_eveqn_q} \\
  \Omega &= 1 - \Sigma^{2} - N_{-}^{2} - 4 A^{2}, \label{eq_eveqn_Omega} \\
  \Sigma^{2} &= \Sigma_{+}^{2} + \Sigma_{-}^{2} + \Sigma_{2}^{2}
    + \Sigma_{\times}^{2} , \label{eq_eveqn_shear}
\end{eqnarray}
and ${}'$ denotes differentiation with respect to $\tau$. There are two
auxiliary equations, the evolution equation for $\Omega$, which reads
\begin{equation}
  \Omega ' = [ 2q - (3\gamma-2) ] \Omega , \label{eq_eveqn_Omega_eveqn}
\end{equation}
and the evolution equation for the constraint function $g(\vec{x})$, namely,
\begin{equation}
  g' = 2(q + \Sigma_{+} - 1) g, \label{eq_eveqn_g}
\end{equation}
which guarantees that $g(\vec{x}) = 0$ defines an invariant set. Both
\eref{eq_eveqn_Omega_eveqn} and \eref{eq_eveqn_g} follow from
\eref{eq_eveqn_eveqn}.

The physical state space is the subset of $\mathbb{R}^{6}$ defined by the
constraint~\eref{eq_eveqn_cons} and the requirement
\begin{equation}
  \Omega \geq 0 . \label{eq_eveqn_statespaceA}
\end{equation}
The restriction \eref{eq_eveqn_statespaceA} implies that the state space is
bounded. In addition, the evolution equations are invariant under the
symmetries
\begin{equation}
  (\Sigma_{+},\Sigma_{-},\Sigma_{2},\Sigma_{\times},N_{-},A) \longrightarrow
  (\Sigma_{+}, \Sigma_{-}, \pm\Sigma_{2}, \pm\Sigma_{\times}, \pm N_{-},
    \pm A ), \label{eq_eveqn_symmetries}
\end{equation}
provided that the product $\Sigma_{\times} N_{-} A$ does not change sign.
Since the evolution equations for $\Sigma_{2}$ and $A$ do not allow
either of these variables to change sign along an orbit, we can assume
without loss of generality that
\begin{equation}
  \Sigma_{2} \geq 0, \qquad A \geq 0 . \label{eq_eveqn_statespaceB}
\end{equation}
Thus the physical state space $\mathcal{D}$ is the subset
$\mathcal{D} \subset \mathbb{R}^{6}$ defined by \eref{eq_eveqn_cons},
\eref{eq_eveqn_statespaceA} and \eref{eq_eveqn_statespaceB}.

The state space $\mathcal{D}$ contains a number of invariant subsets which
play an important role as regards the dynamics in the asymptotic regimes.
Firstly there is the \textit{vacuum boundary}, given by
$\Omega = 0$. Secondly there are various invariant sets with $A = 0$ that
describe vacuum Bianchi I (Kasner) and vacuum Bianchi II (Taub) models, which
we shall introduce in \sref{sec_past}. Thirdly, it follows from the evolution
equations~\eref{eq_eveqn_eveqn} that the conditions\footnote[1]{On account of
the constraint~\eref{eq_eveqn_cons} the first two restrictions imply that
$\Sigma_{+} + \sqrt{3}\, \Sigma_{-} = 0$, since $A \not= 0$ for the
exceptional Bianchi cosmologies.}
\begin{equation}
  \Sigma_{\times} = 0, \qquad N_{-} = 0, \qquad
  \Sigma_{+} + \sqrt{3}\, \Sigma_{-} = 0, \label{eq_eveqn_special_case}
\end{equation}
define a 3-dimensional invariant set, which we shall denote by
$S$. This set describes the subclass of the exceptional Bianchi
cosmologies for which the $G_{2}$ admits one hypersurface-orthogonal Killing
vector field. The asymptotic dynamics of this subclass have been analyzed in
detail by Hewitt (1991). Finally, we note that the invariant set
$\Sigma_{2} = 0$ describes the non-exceptional Bianchi VI$_{-1/9}$ models. In
other words, \textit{the variable $\Sigma_{2}$ describes the
additional shear degree of freedom referred to in the introduction}.

%%%%%%%%%%%%%%%%%%%%%%%%%%%%%%%%%%%%%%%%%%%%%%%%%%%%%%%%%%%%%%%%%%%%%%%%%%%

\section{Local stability of the equilibrium points} \label{sec_eqp}

In this section we discuss the local stability of the equilibrium points of
the DE~\eref{eq_eveqn_eveqn} subject to the constraint~\eref{eq_eveqn_cons}.
The equilibrium points can be found by solving the system of algebraic
equations
\begin{displaymath}
  \vec{f}(\vec{a}) = \mathbf{0}, \qquad g(\vec{a}) = 0.
\end{displaymath}  
The local stability is determined by linearizing the DE~\eref{eq_eveqn_eveqn}
at $\vec{x} = \vec{a}$, which gives
\begin{displaymath}
  \vec{x} ' = \mathrm{D} \vec{f}(\vec{a}) \vec{x},
\end{displaymath}
and finding the eigenvalues of the derivative matrix $\mathrm{D} 
\vec{f}(\vec{a})$. The existence of the constraint~\eref{eq_eveqn_cons}
complicates the analysis, which requires that we consider only the physical
eigenvectors of $\mathrm{D} \vec{f}(\vec{a})$, that is, those which are
tangent to the constraint surface, or equivalently, those which are orthogonal
to the gradient vector $\nabla g(\vec{a})$. Eigenvalues and eigenvectors
that satisfy this condition shall be referred to as \textit{physical}.
We note that if all the physical eigenvalues have negative (positive) real
parts then the equilibrium point is a local sink (source), that is, it attracts
(repels) all orbits in a neighbourhood. In addition to isolated equilibrium
points, we will also encounter arcs of equilibrium points, for which one
physical eigenvalue is necessarily zero. In this case, the criterion for a
local sink (source) is that all eigenvalues other than the zero one have
negative (positive) real parts.

We now list the equilibrium points of the DE~\eref{eq_eveqn_eveqn} subject to
the constraint~\eref{eq_eveqn_cons}, and give the values of the density
parameter $\Omega$ and the deceleration parameter $q$. We refer to WE (see
pp~189--93) for the corresponding exact solutions.

\subsection*{Non-vacuum equilibrium points ($\Omega > 0$)}

\subsection{Flat Friedmann-Lema\^{\i}tre equilibrium point, $\mathsf{FL}$}

\begin{eqnarray*}
  \Sigma_{+} = \Sigma_{-} = \Sigma_{2} = \Sigma_{\times} = N_{-} = A = 0 , \\
  \Omega = 1, \qquad q = \case{1}{2}(3\gamma-2), \qquad 0 < \gamma < 2.
\end{eqnarray*}

\subsection{Collins-Stewart Bianchi II equilibrium points, $\mathsf{CS^{\pm}}$}

\begin{eqnarray*}
  \Sigma_{+} = -\case{1}{16}(3\gamma-2), \qquad
    \Sigma_{-} = \sqrt{3}\, \Sigma_{+}, \qquad N_{-} = \pm \case{1}{8}
    \sqrt{3(3\gamma-2)(2-\gamma)}, \\
    \Sigma_{2} = \Sigma_{\times} = A = 0, \\
  \Omega = \case{3}{16}(6-\gamma), \qquad q = \case{1}{2}(3\gamma-2),
    \qquad \case{2}{3} < \gamma < 2.
\end{eqnarray*} 

\subsection{Collins Bianchi VI$_{-1/9}$ equilibrium point, $\mathsf{C}$}

\begin{eqnarray*}
  \Sigma_{+} = -\case{1}{4}(3\gamma-2), \qquad
  \Sigma_{-} = -\case{1}{\sqrt{3}}\, \Sigma_{+}, \qquad
  A = \case{1}{4} \sqrt{(3\gamma-2)(2-\gamma)}, \\
  \Sigma_{2} = \Sigma_{\times} = N_{-} = 0, \\
  \Omega = \case{1}{3}(5-3\gamma), \qquad q = \case{1}{2}(3\gamma-2), \qquad
  \case{2}{3} < \gamma < \case{5}{3}.
\end{eqnarray*}

\subsection{Wainwright Bianchi VI$_{-1/9}$ arc of equilibrium points,
  $\mathsf{W}$}

\begin{eqnarray*}
  \Sigma_{+} = -\case{1}{3}, \qquad \Sigma_{-} = \case{1}{3\sqrt{3}}, \qquad
  \Sigma_{2} = \case{\sqrt{5}\,\alpha}{3\sqrt{3}}, \\
  \Sigma_{\times} = N_{-} = 0, \qquad A = \sqrt{\case{1}{54}(4+5\alpha^{2})},
  \qquad 0 < \alpha < 1, \\
  \Omega = \case{5}{9}(1-\alpha^{2}), \qquad q = \case{2}{3}, \qquad
  \gamma = \case{10}{9} .
\end{eqnarray*}

\subsection*{Vacuum equilibrium points ($\Omega = 0$)}

\subsection{Robinson-Trautman Bianchi VI$_{-1/9}$ equilibrium point,
  $\mathsf{RT}$}

\begin{eqnarray*}
  \Sigma_{+} = -\case{1}{3}, \qquad \Sigma_{-} = \case{1}{3\sqrt{3}}, \qquad
  \Sigma_{2} = \case{\sqrt{5}}{3\sqrt{3}}, \\
  \Sigma_{\times} = N_{-} = 0, \qquad A = \case{1}{\sqrt{6}}, \\
  q = \case{2}{3} .
\end{eqnarray*}

\subsection{Plane wave arcs of equilibrium points, $\mathsf{PW^{\pm}}$}

\begin{eqnarray*}
  \Sigma_{+} = -\case{1}{4}(4-\alpha), \qquad \Sigma_{-} = \case{1}{4}
    \sqrt{3}\, \alpha, \qquad \Sigma_{2} = 0, \\
  \Sigma_{\times} = \pm \case{1}{2} \sqrt{\alpha(1-\alpha)}, \qquad
  N_{-} = - \Sigma_{\times}, \qquad A = \case{1}{4} \alpha, \qquad
  0 < \alpha \leq 1, \\
  q = \case{1}{2}(4-\alpha) .
\end{eqnarray*}

\subsection{Kasner circle of equilibrium points, $\mathcal{K}$}

\begin{eqnarray*}
  \Sigma_{2} = \Sigma_{\times} = N_{-} = A = 0, \qquad
  \Sigma_{+}^{2} + \Sigma_{-}^{2} = 1, \\
  q = 2 .
\end{eqnarray*}

We now consider whether any of the equilibrium points are local sinks or
sources. It turns out that for each value of the equation of state parameter
$\gamma$ in the interval $0 < \gamma < 2$, there is a unique equilibrium
point/set that is a local sink, as indicated in \tref{table_local_sinks}. We
can describe the transitions, i.e.\ bifurcations, that occur between these
local sinks as follows:
\begin{equation}
  \mathsf{FL} \:\overset{\gamma = \frac{2}{3}}{\longrightarrow}\:
  \mathsf{C} \:\overset{\gamma = \frac{10}{9}}{\longrightarrow}\:
  \mathsf{W} \:\overset{\gamma = \frac{10}{9}}{\longrightarrow}\:
  \mathsf{RT} \label{eq_eqp_bifurcations}
\end{equation}

The mechanisms for these bifurcations, without giving full details, can
be described as follows. The linearization of the evolution equations for
$N_{-}$ and $A$ at the flat $\mathsf{FL}$ point is
\begin{eqnarray*}
  N'_{-} &= \case{1}{2}(3\gamma-2) N_{-}, \qquad
  A' &= \case{1}{2}(3\gamma-2) A,
\end{eqnarray*}
establishing that the spatial curvature variables $N_{-}$ and $A$ destabilize
$\mathsf{FL}$ at $\gamma = \frac{2}{3}$. Moreover when $\gamma = \frac{2}{3}$,
both the Collins-Stewart $\mathsf{CS}$ and Collins $\mathsf{C}$ equilibrium
points bifurcate from the $\mathsf{FL}$ equilibrium point. The linearization
of the evolution equation for $\Sigma_{2}$ at the Collins equilibrium point
$\mathsf{C}$ is
\begin{displaymath}
  \Sigma'_{2} = \case{1}{2} (9\gamma-10) \Sigma_{2},
\end{displaymath}
showing that $\Sigma_{2}$ destabilizes $\mathsf{C}$ at
$\gamma = \frac{10}{9}$. Indeed, at $\gamma = \frac{10}{9}$ there is a line
bifurcation which is characterized by the exchange of stability between the
Collins equilibrium point $\mathsf{C}$ and the Robinson-Trautman vacuum
equilibrium point $\mathsf{RT}$ by means of the Wainwright arc of equilibria
$\mathsf{W}$ which connects both points. It is also of interest to note that
for $\frac{10}{9} < \gamma < \frac{5}{3}$ $\mathsf{C}$ is a saddle, and it
merges with the plane wave arcs (at $\alpha = 1$) when $\gamma = \frac{5}{3}$.

\begin{table}
  \renewcommand{\arraystretch}{1.2}
  \caption{\label{table_local_sinks}Local sinks in the exceptional Bianchi
    state space.}
  \begin{indented} \item[] \begin{tabular}{@{}lllll} \br
  Range of $\gamma$ & Local sink & $\Omega$ & $\Sigma^{2}$ &
    $\mathcal{W}$ \\\mr
  $0 < \gamma \leq \frac{2}{3}$ & $\mathsf{FL}$ & 1 & 0 & 0 \\
  $\frac{2}{3} < \gamma < \frac{10}{9}$ & $\mathsf{C}$ &
    $\frac{1}{3}(5-3\gamma)$ & $\frac{1}{12}(3\gamma-2)^{2}$ &
    $\frac{1}{6}(3\gamma-2)\sqrt{3\gamma+1}$ \\
  $\gamma = \frac{10}{9}$ & $\mathsf{W}$ &
    $\frac{5}{9}(1-\alpha^{2})$ & $\frac{1}{27}(4+5\alpha^{2})$ &
    $\frac{1}{9\sqrt{3}}\sqrt{(4+5\alpha^2)(13+35\alpha^2)}$ \\
  $\frac{10}{9} < \gamma < 2$ & $\mathsf{RT}$ & 0 &
    $\frac{1}{3}$ & $\frac{4}{3}$ \\ \br
  \end{tabular}\end{indented}
\end{table}

The local sinks in the exceptional Bianchi state space are listed in
\tref{table_local_sinks}, together with the values of the density parameter
$\Omega$, the shear parameter $\Sigma$, which describes the anisotropy in the
fluid congruence (see WE, p~114) and the Weyl curvature parameter
$\mathcal{W}$, which can be regarded as a measure of the intrinsic anisotropy
of the gravitational field (see Wainwright \etal 1999, p~2580). For the present
class of models, $\Omega$ and $\Sigma^{2}$ are given by \eref{eq_eveqn_Omega}
and \eref{eq_eveqn_shear}. The formula for $\mathcal{W}$ is provided in
appendix~B.

It is of interest to note that the plane wave arcs of equilibrium points,
$\mathsf{PW^{\pm}}$, which describe the SH plane wave vacuum solutions,
do not appear in the table. This result represents an important difference
between the exceptional and non-exceptional Bianchi VI$_{h}$ models. For the
latter class, the plane wave equilibrium points are a local sink for vacuum
models and for perfect fluid models, subject to a restriction on the equation
of state parameter $\gamma$. On the other hand, the plane wave arcs of
equilibria are saddles in the exceptional Bianchi state space. To see this, we
linearize the evolution equation for $\Sigma_{2}$ at the plane wave arcs
obtaining
\begin{displaymath}
  \Sigma'_{2} = \case{1}{2}(6-\alpha) \Sigma_{2}.
\end{displaymath}
Thus, \textit{the extra shear degree of freedom, $\Sigma_{2}$, that is present
in the exceptional Bianchi cosmologies, destabilizes the plane wave solutions}.
Their stability is, in effect, inherited by the Robinson-Trautman solution.

As regards local sources, it follows upon analysis of the eigenvalues
associated with the equilibrium points that \textit{there are no local sources
in the exceptional Bianchi state space}. In particular, the Kasner circle
$\mathcal{K}$ is a saddle, and, consequently, the initial state of a typical
model cannot be a single Kasner equilibrium point.

%%%%%%%%%%%%%%%%%%%%%%%%%%%%%%%%%%%%%%%%%%%%%%%%%%%%%%%%%%%%%%%%%%%%%%%%%%%

\section{The late-time asymptotic regime} \label{sec_late}

We have seen that for each value of $\gamma$ in the range $0 < \gamma < 2$,
excluding $\gamma = \frac{10}{9}$, there is a unique equilibrium point that is
a local sink of the evolution equations, while if $\gamma = \frac{10}{9}$,
the local sink is an arc of equilibrium points. We note that all of the local
sinks in \tref{table_local_sinks} lie in the three-dimensional invariant set
$S$ defined by \eref{eq_eveqn_special_case}. It has been proved that each of
these local sinks is the future attractor in the invariant set $S$, for the
relevant range of $\gamma$ (see Hewitt~1991)\footnote[1]{The analysis in this
paper is complete, except for the case $\gamma = \frac{10}{9}$. We have
recently been able to complete this analysis by finding a Dulac function for
the family of two-dimensional invariant subsets that foliate the state space.
We thank Neville Dubash for assistance with this matter.}.

We have performed extensive numerical experiments which suggest that the 
solution determined by an arbitrary initial condition satisfies
\begin{equation}
  \lim_{\tau \rightarrow +\infty} \Sigma_{\times} = 0, \qquad
  \lim_{\tau \rightarrow +\infty} N_{-} = 0, \qquad
  \lim_{\tau \rightarrow +\infty} \Sigma_{+} + \sqrt{3}\, \Sigma_{-} = 0.
    \label{eq_late_limits}
\end{equation}
In other words, the experiments provide strong evidence that all orbits are
attracted to the invariant set $S$, i.e.\ that the future attractor is
contained in this subset. We thus make the following conjecture.

\begin{conjecture}
  Each local sink given in \tref{table_local_sinks} for $\gamma$ in the
  range $0 < \gamma < 2$ is the future attractor in the exceptional Bianchi
  state space. \label{conjecture_future}
\end{conjecture}

\noindent This conjecture is in fact known to be true for $\gamma$ in the range
$0 < \gamma < \frac{2}{3}$ (see WE, theorem 8.2, p~174). Giving a proof for
$\gamma$ in the range $\frac{2}{3} \leq \gamma < 2$ necessitates establishing
the limits \eref{eq_late_limits}.

%%%%%%%%%%%%%%%%%%%%%%%%%%%%%%%%%%%%%%%%%%%%%%%%%%%%%%%%%%%%%%%%%%%%%%%%%%%

\section{The singular asymptotic regime} \label{sec_past}

In this section we show that the exceptional Bianchi cosmologies exhibit an
oscillatory approach to the initial singularity, as do the Bianchi VIII and IX
models. There are two significant differences, however, which we now describe.

In the dynamical systems approach, the mechanism for creating an oscillatory
singularity is that the Kasner equilibrium points, which comprise the Kasner
circle, are saddles, and the unstable manifold (into the past) of any Kasner
point is asymptotic to another Kasner point. In other words, the unstable
manifold consists of orbits that join two Kasner points. We shall refer to
these unstable manifolds as \textit{Kasner transition sets}, because they
provide a mechanism for a cosmological model to make a transition from one
(approximate) Kasner state to another, as it evolves into the past. For
non-tilted Bianchi VIII and IX models, each Kasner transition set is
\textit{one-dimensional} and thus there is a unique heteroclinic
orbit\footnote[1]{A heteroclinic orbit is one that joins two equilibrium
points.} joining two Kasner points. In the present case, for two of the six
arcs on the Kasner circle, the Kasner transition set is
\textit{two-dimensional}, leading to a one-parameter family of heteroclinic
orbits. Nevertheless, in both cases, the stability properties of the Kasner
equilibrium points lead to the existence of infinite heteroclinic sequences,
i.e.\ infinite sequences of equilibrium points on the Kasner circle
joined by special heteroclinic orbits, oriented into the past (see WE,
section~6.4.2, for Bianchi VIII and IX models). These heteroclinic sequences
govern the dynamics in the singular regime ($\tau \rightarrow -\infty$) in the
sense that a typical orbit shadows (i.e.\ is approximated by) a heteroclinic
sequence as $\tau \rightarrow -\infty$. In physical terms, the corresponding
cosmological model is approximated by a sequence of Kasner vacuum models as
the singularity is approached into the past, the so-called \textit{Mixmaster
oscillatory singularity}.

\begin{figure}
  \begin{indented} \item[] \includegraphics[scale=0.6]{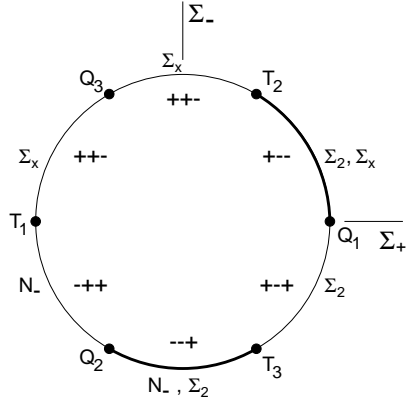}
    \end{indented}
  \caption{\label{fig1}The arrays $++-$, etc.\ give the signs of the eigenvalues
  $\lambda_{N_{-}}$, $\lambda_{\Sigma_{2}}$ and $\lambda_{\Sigma_{\times}}$
  in that order. The variables(s) listed next to each of the six arcs
  indicates which of the variable(s) $N_{-}$, $\Sigma_{2}$, $\Sigma_{\times}$
  is growing into the past. Kasner points on the bold arcs $T_{2}Q_{1}$ and
  $T_{3}Q_{2}$ have a two-dimensional Kasner transition set into the past.}
\end{figure}

The second difference is that in the case of Bianchi VIII and IX models, the
instability of the Kasner points is generated by spatial curvature only,
while in the case of exceptional Bianchi cosmologies, the instability is
due to a combination of spatial curvature and off-diagonal shear, in
particular, the variables $N_{-}$, $\Sigma_{2}$ and $\Sigma_{\times}$.

Figure~\ref{fig1} shows the signs of the eigenvalues of the Kasner equilibrium
points associated with the variables $N_{-}$, $\Sigma_{2}$ and
$\Sigma_{\times}$, showing that points on the arcs $T_{2}Q_{1}$ and
$T_{3}Q_{2}$ have a two-dimensional Kasner transition set into the past
(negative eigenvalues indicate instability into the past). The figure also
shows which of the three variables are increasing into the past in a
neighbourhood of the Kasner circle.

Figure~\ref{fig2} shows the projections in the $\Sigma_{+}\Sigma_{-}$-plane,
of three families of heteroclinic orbits, that join two Kasner points. For a
Kasner point with a one-dimensional transition set, the orbit shown
in \fref{fig2} emanating from that point is the transition set. These
families are described as follows.

\subsection*{$\mathcal{S}_{N_{-}}$: Vacuum Bianchi II models (Taub models)}

These orbits describe the familiar vacuum Bianchi II models. They are given by
\begin{equation}
  \Sigma_{2} = \Sigma_{\times} = A = 0, \qquad \Sigma_{+}^{2} + \Sigma_{-}^{2}
    + N_{-}^{2} = 1, \qquad N_{-} \not= 0, \label{eq_past_S_Nm}
\end{equation}
with $\Sigma_{-} + \sqrt{3} = m(\Sigma_{+} + 1)$, where
$m > \frac{1}{\sqrt{3}}$ is a constant.

\subsection*{$\mathcal{S}_{\Sigma_{\times}}$: Kasner models relative to a
non-Fermi-propagated frame, with $\Sigma_{2} = 0$}

These orbits are given by
\begin{equation}
  \Sigma_{2} = N_{-} = A = 0, \qquad \Sigma_{+}^{2} + \Sigma_{-}^{2}
    + \Sigma_{\times}^{2} = 1, \qquad \Sigma_{\times} \not= 0,
    \label{eq_past_S_Sc}
\end{equation}
with $\Sigma_{+} = m$, where $-1 < m < 1$ is a constant.

\subsection*{$\mathcal{S}_{\Sigma_{2}}$: Kasner models relative to a
non-Fermi-propagated frame, with $\Sigma_{\times} = 0$}

These orbits are given by
\begin{equation}
  \Sigma_{\times} = N_{-} = A = 0, \qquad \Sigma_{+}^{2} + \Sigma_{-}^{2}
    + \Sigma_{2}^{2} = 1, \qquad \Sigma_{2} > 0, \label{eq_past_S_S2}
\end{equation}
with $\Sigma_{+} + \sqrt{3}\, \Sigma_{-} = m$, where $-2 < m < 2$ is a
constant.\vspace{2ex plus 0.4ex minus 0.4ex}

\begin{figure}
  \renewcommand{\mathindent}{0in}
  \begin{tabular}{@{}ccc}
    \includegraphics[scale=0.46]{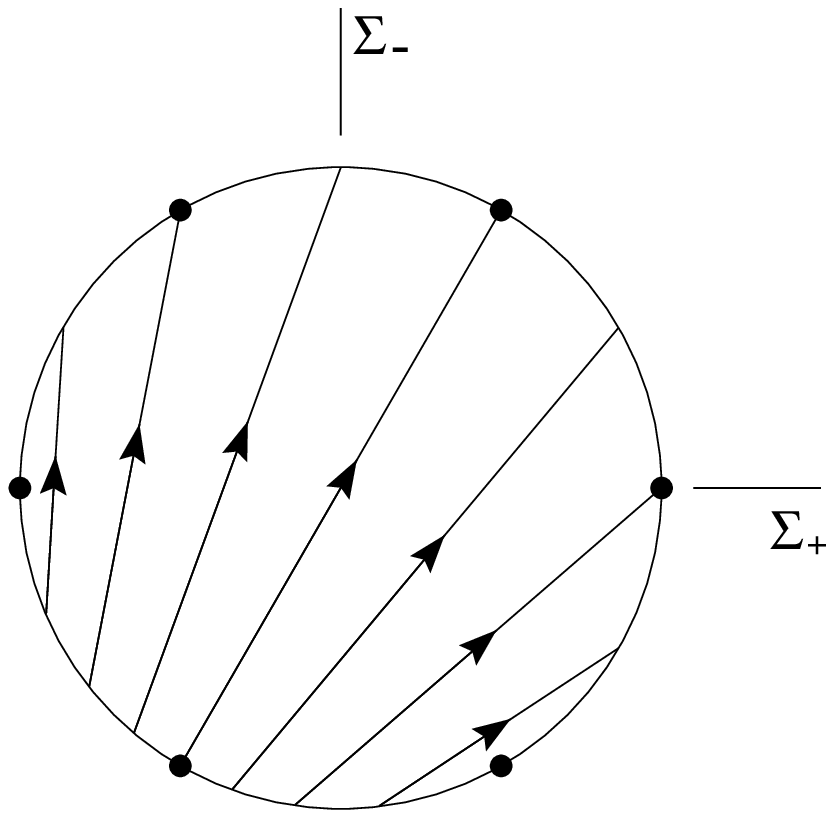} &
    \includegraphics[scale=0.46]{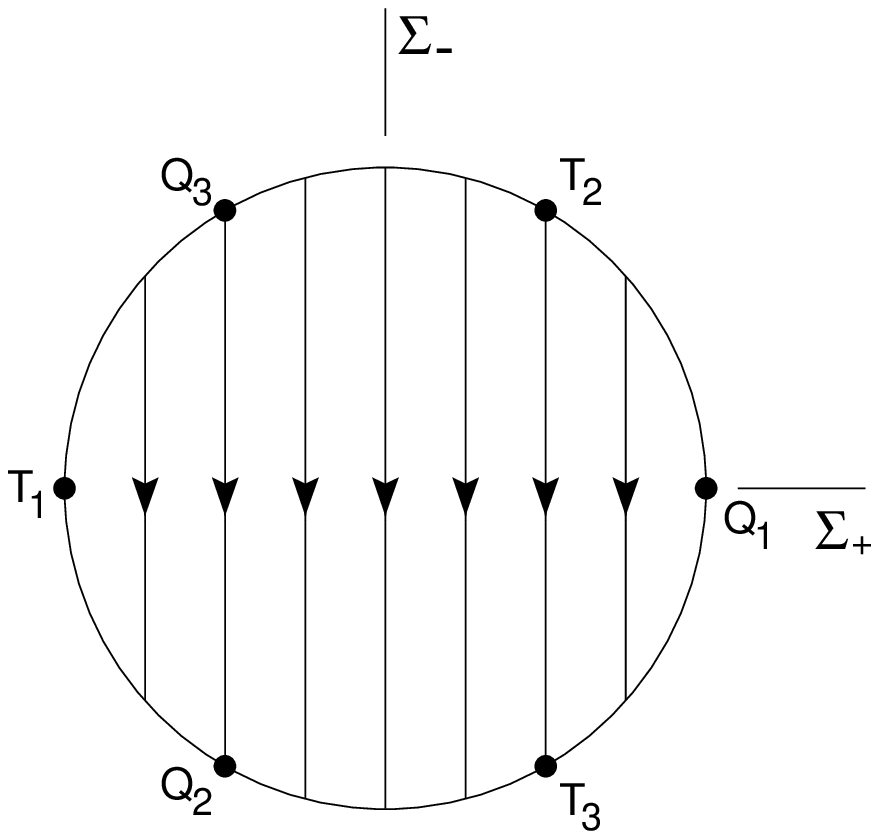} &
    \includegraphics[scale=0.46]{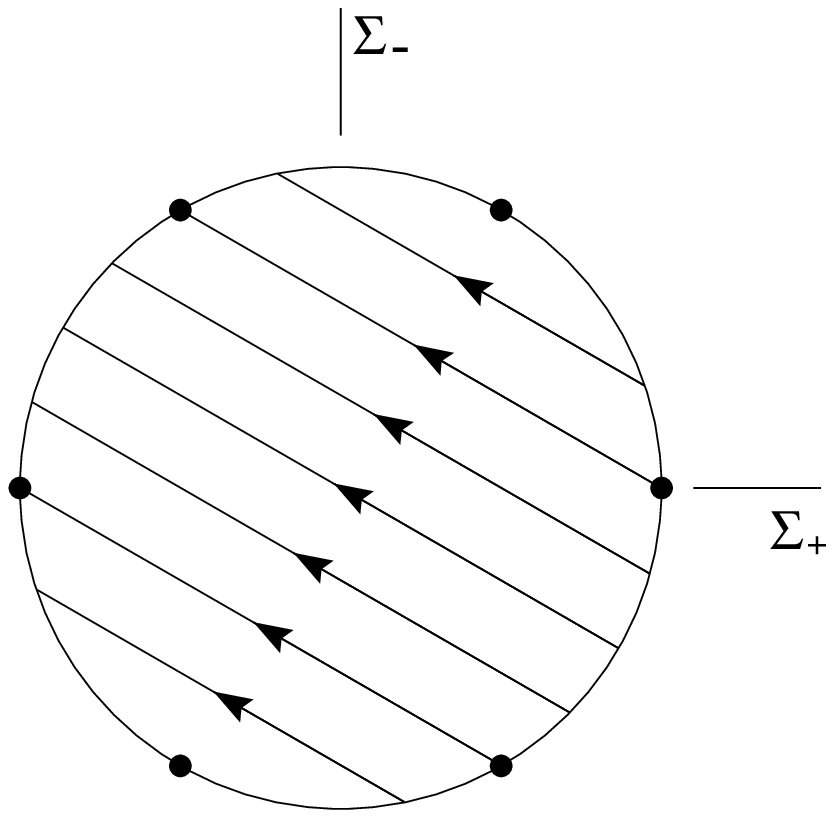} \\
    \footnotesize i) $N_{-} \not = 0$ &
    \footnotesize ii) $\Sigma_{\times} \not= 0$ &
    \footnotesize iii) $\Sigma_{2} > 0$
  \end{tabular}
  \caption{\label{fig2}Vacuum orbits joining points on the Kasner circle
    $\mathcal{K}$. The arrows show evolution into the past.}
\end{figure}

\begin{figure}
  \renewcommand{\mathindent}{0in}
  \begin{minipage}[t]{0.49\textwidth}
    \includegraphics[scale=0.7]{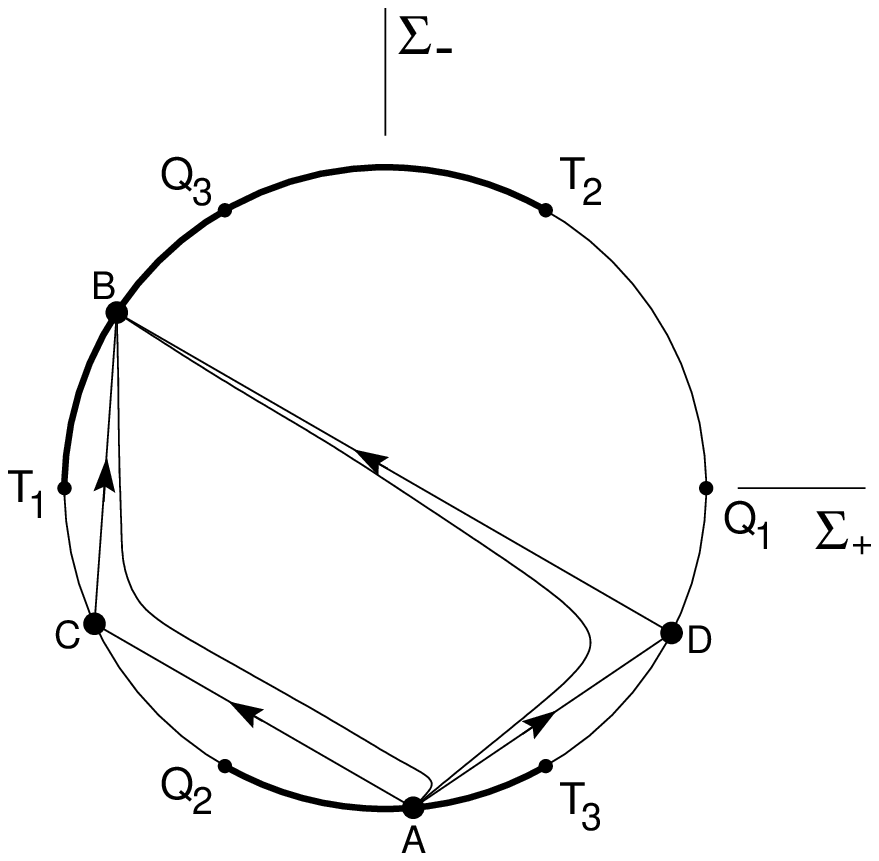}
    \caption{\label{fig3}The two-dimensional Kasner transition set of a
      Kasner point $A$ on the arc $T_{3}Q_{2}$, projected onto the
      $\Sigma_{+}\Sigma_{-}$-plane. The arrows show evolution into the past.}
  \end{minipage}\hfill
  \begin{minipage}[t]{0.49\textwidth}
    \includegraphics[scale=0.7]{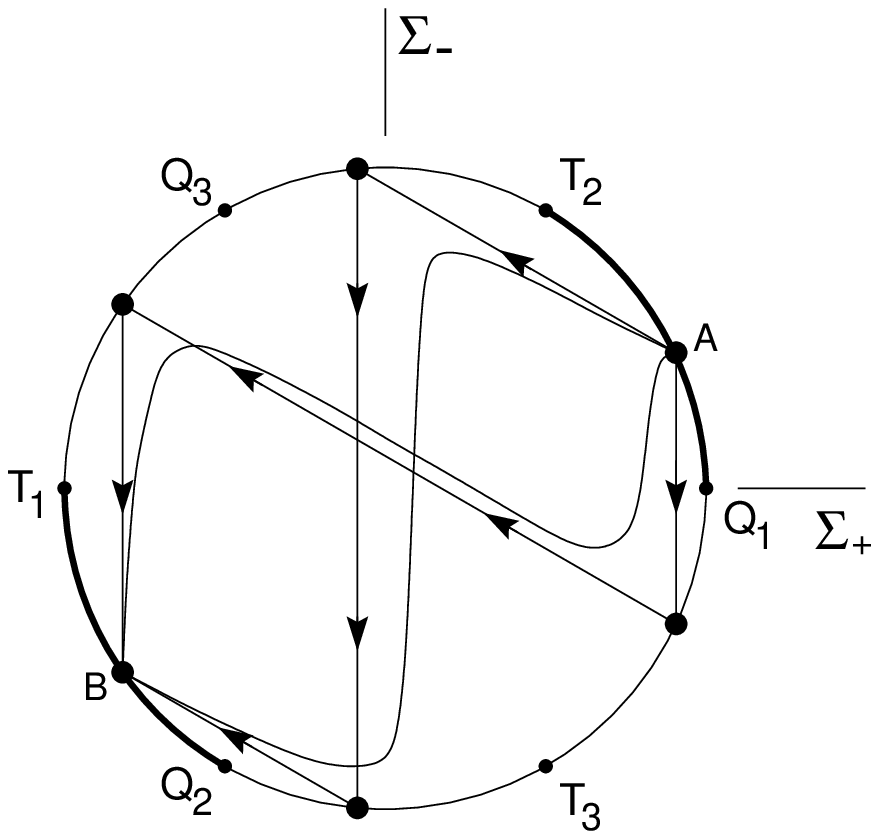}
    \caption{\label{fig4}The two-dimensional Kasner transition set of a
      Kasner point $A$ on the arc $T_{2}Q_{1}$, projected onto the
     $\Sigma_{+}\Sigma_{-}$-plane. The arrows show evolution into the past.}
  \end{minipage}
\end{figure}

Figure~\ref{fig3} gives a representation of the two-dimensional transition set
of a Kasner point on the arc $T_{3}Q_{2}$. This invariant set consists of a
one-parameter family of orbits with $\Omega = 0$, $A = 0$,
$\Sigma_{\times} = 0$, $N_{-} \not= 0$ and $\Sigma_{\times} \not= 0$, which
join $A$ to $B$ (the projection of two typical orbits is shown), and two special
orbits, one with $N_{-} \not= 0$ and $\Sigma_{2} = \Sigma_{\times} = 0$
(a member of the family $\mathcal{S}_{N_{-}}$), and the other with
$\Sigma_{2} \not= 0$ and $N_{-} = \Sigma_{\times} = 0$ (a member of the family
$\mathcal{S}_{\Sigma_{2}}$). These orbits, which describe the vacuum Bianchi
II models relative to a non-Fermi-propagated frame, can be described
explicitly by the equation\footnote[1]{This expression is a first
integral of the evolution equations \eref{eq_eveqn_eveqn} when restricted to
the invariant set $\Omega = 0$, $A = 0$, $\Sigma_{\times} = 0$. An
equivalent expression has arisen in the analysis of $G_{2}$ cosmologies
(see Weaver 2002). We thank Marsha Weaver for pointing out this first integral
in the $G_{2}$ context, and Woei~Chet Lim for calculating the expression in
\eref{eq_past_firstintegral_A}.}
\begin{equation}
  \frac{ (1+\Sigma_{+}+\sqrt{3}\,\Sigma_{-})^{2} + 3 N_{-}^{2} }{
    (4+\Sigma_{+}+\sqrt{3}\,\Sigma_{-})^{2}} = \alpha,
    \label{eq_past_firstintegral_A}
\end{equation}
where $\alpha$ is a constant that depends on the point $A$.

Figure~\ref{fig4} gives a representation of the two-dimensional transition set
of a Kasner point on the arc $T_{2}Q_{1}$. This invariant set consists of a
one-parameter family of orbits with $\Omega = 0$, $A = 0$, $N_{-} = 0$,
$\Sigma_{2} \not= 0$ and $\Sigma_{\times} \not= 0$, which join $A$ to $B$
(the projection of two typical orbits is shown), and two special orbits, one
with $\Sigma_{2} \not= 0$ and $\Sigma_{\times} = N_{-} = 0$ (a member of the
family $\mathcal{S}_{\Sigma_{2}}$), and the other with
$\Sigma_{\times} \not= 0$ and $N_{-} = \Sigma_{2} = 0$ (a member of the
family $\mathcal{S}_{\Sigma_{\times}}$). All orbits in this manifold, which
intersect the Kasner circle in six points, as shown in \fref{fig4}, represent
one and the same physical Kasner model. The six Kasner points referred to have
the same Kasner exponents, but differ in the labelling, and the orbits in the
Kasner transition set thus describe a rotation of the spatial axes relative to
which this Kasner model is described. These orbits can be described explicitly
by the equation\footnote[2]{This expression is a first integral of the
evolution equations \eref{eq_eveqn_eveqn} when restricted to the invariant set
$\Omega = 0$, $A = 0$, $N_{-} = 0$. It is related to a Hubble-normalized
scalar formed from the Weyl tensor, which is constant for Kasner models. We
thank Woei~Chet Lim for verifying this result.}
\begin{equation}
  2 - 6 \Sigma_{+} + 8 \Sigma_{+}^{3} + 3\sqrt{3}( \sqrt{3}\,\Sigma_{+}
    + \Sigma_{-}) \Sigma_{2}^{2} = \alpha, \label{eq_past_firstintegral_B}
\end{equation}
where $\alpha$ is a constant that depends on the point $A$, thereby
distinguishing different Kasner models.

Numerical experiments suggest that as $\tau \rightarrow -\infty$, an orbit
that is close to a point $A$ on the arc $T_{3}Q_{2}$ will shadow either the
orbit $AD \subset \mathcal{S}_{N_{-}}$ or the orbit $AC \subset
\mathcal{S}_{\Sigma_{2}}$, rather than shadowing a typical orbit in the
two-dimensional Kasner transition set. A similar remark applies to points on
the arc $T_{2}Q_{1}$. The result of this behaviour is that as
$\tau \rightarrow -\infty$ an orbit will shadow a heteroclinic sequence that
consists of orbits belonging to the invariant sets $\mathcal{S}_{N_{-}}$,
$\mathcal{S}_{\Sigma_{2}}$ and $\mathcal{S}_{\Sigma_{\times}}$. This
behaviour motivates the following conjecture concerning the past attractor
$\mathcal{A}^{-}$.

\begin{conjecture}
  The past attractor is the two-dimensional invariant set consisting of all
  orbits in the invariant sets $\mathcal{S}_{N_{-}}$,
  $\mathcal{S}_{\Sigma_{\times}}$ and $\mathcal{S}_{\Sigma_{2}}$ (see
  \fref{fig2}) and the Kasner equilibrium points, i.e.
  \begin{equation}
    \mathcal{A}^{-} = \mathcal{S}_{N_{-}} \cup \mathcal{S}_{\Sigma_{\times}}
      \cup \mathcal{S}_{\Sigma_{2}} \cup \mathcal{K} .
      \label{eq_past_pastattractor}
  \end{equation} \label{conjecture_past}
\end{conjecture}

This conjecture can be formulated in terms of limits of the state variables
as follows. Referring to \eref{eq_past_S_Nm}--\eref{eq_past_S_S2},
\eref{eq_eveqn_Omega} and \eref{eq_eveqn_shear} we see that the set
$\mathcal{A}^{-}$ is defined by
\begin{displaymath}
  \Omega = 0, \qquad A = 0
\end{displaymath}
and
\begin{displaymath}
  N_{-} \Sigma_{2} = 0, \qquad \Sigma_{2} \Sigma_{\times} = 0, \qquad
  \Sigma_{\times} N_{-} = 0,
\end{displaymath}
the final equality being a consequence of $A = 0$ and the constraint
\eref{eq_eveqn_cons}. Thus our conjecture concerning the past attractor
can be formulated as
\begin{displaymath}
  \lim_{\tau \rightarrow -\infty} \Omega = 0, \qquad
  \lim_{\tau \rightarrow -\infty} \Delta = 0,
\end{displaymath}
where
\begin{displaymath}
  \Delta = (N_{-}\Sigma_{2})^{2} + (\Sigma_{2}\Sigma_{\times})^{2}
    + (\Sigma_{\times}N_{-})^{2}.
\end{displaymath}
These limits imply that $\lim_{\tau \rightarrow -\infty} A = 0$, on account
of \eref{eq_eveqn_cons}. Note that for a generic orbit, $\lim_{\tau
\rightarrow -\infty} (N_{-},\Sigma_{2},\Sigma_{\times})$ does not exist.

%%%%%%%%%%%%%%%%%%%%%%%%%%%%%%%%%%%%%%%%%%%%%%%%%%%%%%%%%%%%%%%%%%%%%%%%%%%

\section{Discussion} \label{sec_discussion}

There are two generic classes of \textit{non-tilted} ever-expanding SH
cosmologies, namely
\begin{enumerate}
  \item[i)] the Bianchi type VIII cosmologies,
    and\vspace{-0.5ex plus 0.1ex minus 0.1ex}
  \item[ii)] the exceptional Bianchi cosmologies.
\end{enumerate}
The evolution of the first class in the asymptotic regimes has been described
in the literature (see for example, WE, pp~143--7 for the singular regime, and
the accompanying paper, Horwood \etal 2002, for the late-time regime).
In this paper we have used Hubble-normalized variables to describe the
asymptotic dynamics of the second class. We have shown that with this choice
of variables, the Einstein field equations reduce to an autonomous DE on a
5-dimensional compact subset of $\mathbb{R}^{6}$. Since the state space is
compact, there is a past attractor and a future attractor, which we have been
able to identify using heuristic arguments and numerical experiments, thereby
describing the dynamics in the singular and late-time asymptotic regimes.

The mathematical description of the exceptional Bianchi cosmologies differs
from that of the non-exceptional models in the following ways:
\begin{enumerate}
  \item[i)] the expansion-normalized state space is of dimension five rather
    than four, corresponding to the presence of an additional shear degree
    of freedom,
  \item[ii)] the Abelian $G_{2}$ subgroup of the isometry group does not act
    orthogonally transitively.
\end{enumerate}
We have shown that these differences lead to significant changes in the
asymptotic dynamics:
\begin{enumerate}
  \item[i)] the singular regime is oscillatory (infinite sequence of Kasner
    states), rather than asymptotically self-similar,
  \item[ii)] there is a new asymptotic state described by the vacuum SH
    Robinson-Trautman solution, which describes the late-time behaviour of
    vacuum models and of perfect fluid models with $\gamma$ satisfying
    $\frac{10}{9} < \gamma < 2$.
\end{enumerate}
These two points also highlight the similarities and the differences between
the two generic classes of ever-expanding models: both classes have an
oscillatory singularity, but differ as regards the late-time asymptotic
regime. The Bianchi VIII models are not asymptotically self-similar at late
times and exhibit Weyl curvature dominance (the Weyl curvature parameter
$\mathcal{W}$ diverges as $\tau \rightarrow +\infty$, see Horwood \etal 2002),
whereas the exceptional Bianchi cosmologies are asymptotically self-similar
and hence do not exhibit Weyl curvature dominance ($\mathcal{W}$ has a finite
limit as $\tau \rightarrow +\infty$, given in \tref{table_local_sinks}).

We now discuss some aspects of the oscillatory singular regime that occurs in
SH cosmologies. This type of singularity, also known as a Mixmaster
singularity, was first discovered in vacuum Bianchi IX cosmologies (see Misner
1969 and Belinskii \etal 1970) and subsequently in the closely related Bianchi
VIII cosmologies (see for example, WE, section 6.4). Within the dynamical
systems framework, the oscillatory singularity in these models is created by
the three degrees of freedom in the spatial curvature, which destabilize the
Kasner equilibrium points in the Hubble-normalized state space. In the present
models, there is only one degree of freedom in the spatial curvature, but
instead there appear \textit{two off-diagonal shear degrees of freedom}, which
again destabilize the Kasner equilibrium points. This destabilization, and the
resulting existence of the Kasner transition sets (see \sref{sec_past}), leads
to the creation of infinite sequences of heteroclinic orbits joining Kasner
equilibrium points, resulting in an oscillatory singularity. This phenomenon
can in fact occur in models of all Bianchi types, if one permits sufficient
general source terms for the gravitational field. Two classes which have been
analyzed in detail and are closely related to the present class, are the
Bianchi I magnetic cosmologies (see LeBlanc~1997) which have one magnetic
degree of freedom and two off-diagonal shear degrees of freedom, and the
tilted perfect fluid Bianchi II cosmologies (see Hewitt \etal 2001) which have
one spatial curvature degree of freedom and two off-diagonal shear degrees of
freedom. Indeed, the common feature of all SH models with an oscillatory
singularity is the occurrence of the above \textit{Kasner destabilization},
which can be caused by any combination of spatial curvature, off-diagonal
shear or magnetic degrees of freedom totaling at least three.

We conclude by discussing the implications of our results for $G_{2}$
cosmologies. As mentioned in the introduction, the present class of SH
cosmologies are the SH limit of the most general $G_{2}$ cosmologies,
namely those for which the $G_{2}$ does not act orthogonally transitively.
One thus expects that the initial singularity in a general $G_{2}$ model will
be oscillatory, and indeed numerical evidence has been provided that suggests
this is the case, at least for vacuum models (see Berger \etal 2001 and
Lim~2002). Secondly, the fact that the SH Robinson-Trautman solution is a local
sink in the Hubble-normalized state space for vacuum models and for perfect
fluid models with equation of state parameter $\gamma$ satisfying
$\frac{10}{9} < \gamma < 2$ suggests that this solution may play a role in the
dynamics of general $G_{2}$ cosmologies at late times. It would thus be of
interest to investigate the stability of the SH Robinson-Trautman solution in
this context.

%%%%%%%%%%%%%%%%%%%%%%%%%%%%%%%%%%%%%%%%%%%%%%%%%%%%%%%%%%%%%%%%%%%%%%%%%%%

\ack

The authors wish to Woei~Chet Lim, Marsha Weaver and Claes Uggla for helpful
comments and discussions. This research was supported financially by
the Natural Sciences and Engineering Research Council of Canada through
research grants (CH, JW) and an Undergraduate Research Award to JH.

%%%%%%%%%%%%%%%%%%%%%%%%%%%%%%%%%%%%%%%%%%%%%%%%%%%%%%%%%%%%%%%%%%%%%%%%%%%

\appendix
\def\thesection{\Alph{section}}

\stepcounter{section}
\section*{Appendix~\thesection.~Derivation of the evolution equations}

In this appendix we write the EFEs for the exceptional Bianchi cosmologies
as an autonomous system of ordinary differential equations subject to one
algebraic constraint. Our starting point is the system of evolution equations
for a general spatially homogeneous cosmological model given in Hewitt \etal
(2001) (see equations (A.11)--(A.18)). These equations have been obtained by
writing the orthonormal frame equations for a spatially homogeneous model
using a group invariant orthonormal frame $\{\vec{e}_{0},\vec{e}_{\alpha}\}$,
choosing $\vec{e}_{0}$ to be the unit normal to the group orbits. Furthermore,
all commutation functions and source terms are normalized with the Hubble
scalar $H$.

We now describe the specializations that lead to the exceptional Bianchi
cosmologies. Firstly, the models are non-tilted in the sense that the unit
normal $\vec{e}_{0}$ is aligned with the fluid velocity $\vec{u}$. Furthermore,
the models have a perfect fluid source with a $\gamma$-law equation of state
given by
\begin{displaymath}
  P = (\gamma-1)\Omega,
\end{displaymath}
where $P$ is the Hubble-normalized pressure and $\Omega$ is the
dimensionless density parameter. The equation of state parameter $\gamma$ is
restricted to the interval $0 < \gamma < 2$. Secondly, the orthonormal frame
vectors $\vec{e}_{2}$ and $\vec{e}_{3}$ are chosen to be tangential to the
orbits of the two-dimensional Abelian subgroup $G_{2}$ of the
three-dimensional symmetry group $G_{3}$. The consequences of this frame
choice are
\begin{equation}
  N_{1\alpha} = 0, \qquad A_{\alpha} = \delta^{1}{}_{\alpha} A_{1}, \qquad
  R_{3} = -\Sigma_{12}, \qquad R_{2} = \Sigma_{13} \label{eq_app_eveqn_frame1}
\end{equation}
(see Wainwright~1979, pp~2019--21). Equation (A.18) in Hewitt \etal (2001)
contains three constraints, one of which will give rise to the
constraint~\eref{eq_eveqn_cons} and the other two which together lead to the
following possibilities:
\begin{enumerate}
  \item[i)]  $\Sigma_{12} = \Sigma_{13} = 0$,\\
             or
  \item[ii)] $9 A_{1}^{2} - N_{23}^{2} + N_{22} N_{33} = 0 \quad\mathrm{and}
             \quad N_{22} \Sigma_{12} + (N_{23} - 3 A_{1}) \Sigma_{13} = 0$ .
\end{enumerate}
Making the former choice is equivalent to insisting that the Abelian $G_{2}$
subgroup acts orthogonally transitively. We make the latter choice, which
results in the models being Bianchi VI$_{h}$ with\footnote[1]{Recall that the
group parameter $h$ is related to $A_{\alpha}$ and $N_{\alpha\beta}$ according
to\vspace{-1.3ex plus 0.2ex minus 0.2ex}
\begin{displaymath}
  A_{\alpha} A^{\alpha} = \case{1}{2} h \left[ (N_{\alpha}{}^{\alpha})^{2}
    - N_{\alpha}{}^{\beta} N_{\beta}{}^{\alpha} \right].
\end{displaymath}
} $h = -\frac{1}{9}$.

The remaining gauge freedom at this point is an initial alignment in the
$G_{2}$ orbits. We choose to set $\Sigma_{12} = 0$ initially. It follows that
this choice is preserved by the evolution equation for $\Sigma_{12}$ provided
that
\begin{equation}
  R_{1} = \Sigma_{23} . \label{eq_app_eveqn_frame2}
\end{equation}
The gauge freedom in the angular velocity variables $R_{\alpha}$ of the
spatial frame allows us to make such a choice for $R_{1}$. Thus, the
constraints in case ii) now yield
\begin{equation}
  N_{23} - 3 A_{1} = 0, \qquad N_{33} = 0. \label{eq_app_eveqn_frame3}
\end{equation}

At this stage, the independent non-zero Hubble-normalized variables are
\begin{displaymath}
  \Sigma_{22}, \quad \Sigma_{33}, \quad \Sigma_{13}, \quad \Sigma_{23}, \quad
  N_{22}, \quad A_{1}.
\end{displaymath}
We relabel the shear variables according to
\begin{eqnarray}
  \eqalign{
  \Sigma_{+} = \case{1}{2}(\Sigma_{22} + \Sigma_{33}), \qquad
  \Sigma_{-} = \case{1}{2\sqrt{3}}(\Sigma_{22} - \Sigma_{33}), \\
  \Sigma_{2} = \case{1}{\sqrt{3}}\, \Sigma_{13}, \qquad
  \Sigma_{\times} = \case{1}{\sqrt{3}}\, \Sigma_{23}, }
    \label{eq_app_eveqn_frame4}
\end{eqnarray}
and use the notation
\begin{equation}
  N_{-} = \case{1}{2\sqrt{3}}\, N_{22}, \qquad A = A_{1} .
    \label{eq_app_eveqn_frame5}
\end{equation}

The evolution equations~\eref{eq_eveqn_eveqn} and the
constraint~\eref{eq_eveqn_cons} for the exceptional Bianchi cosmologies
now follow from equations (A.10)--(A.18) in Hewitt \etal (2001).

%%%%%%%%%%%%%%%%%%%%%%%%%%%%%%%%%%%%%%%%%%%%%%%%%%%%%%%%%%%%%%%%%%%%%%%%%%%

\stepcounter{section}
\section*{Appendix~\thesection.~The Weyl curvature tensor}

In this appendix we give an expression for the Weyl curvature parameter
$\mathcal{W}$ in terms of the Hubble-normalized variables $\Sigma_{+}$,
$\Sigma_{-}$, $\Sigma_{2}$, $\Sigma_{\times}$, $N_{-}$ and $A$. The Weyl
curvature parameter is defined by
\begin{equation}
  \mathcal{W}^{2} = \frac{ E_{ab} E^{ab} + H_{ab} H^{ab} }{6H^{4}},
    \label{eq_app_Weyl_defn}
\end{equation}
where $E_{ab}$ and $H_{ab}$ are the electric and magnetic parts of the Weyl
tensor, respectively (see WE, p~19), relative to the fluid congruence. Let
$E_{\alpha\beta}$ and $H_{\alpha\beta}$ be the components of $E_{ab}$ and
$H_{ab}$ relative to the group invariant frame introduced in appendix~A. The
Hubble-normalized counterparts of $E_{\alpha\beta}$ and $H_{\alpha\beta}$ are
defined according to
\begin{equation}
  \mathcal{E}_{\alpha\beta} = \frac{E_{\alpha\beta}}{H^{2}}, \qquad
  \mathcal{H}_{\alpha\beta} = \frac{H_{\alpha\beta}}{H^{2}}.
    \label{eq_app_Weyl_a}
\end{equation}
Since $\mathcal{E}_{\alpha\beta}$ and $\mathcal{H}_{\alpha\beta}$ are symmetric
and trace-free, they each have five independent components. We label these in
analogy with the labelling of the shear variables as given in appendix~A, i.e.
\begin{eqnarray}
  \eqalign{
  \mathcal{E}_{+} = \case{1}{2}(\mathcal{E}_{22}+\mathcal{E}_{33}), \qquad
  \mathcal{E}_{-} = \case{1}{2\sqrt{3}}(\mathcal{E}_{22}-\mathcal{E}_{33}), \\
  \mathcal{E}_{\times} = \case{1}{\sqrt{3}}\, \mathcal{E}_{23}, \qquad
  \mathcal{E}_{2} = \case{1}{\sqrt{3}}\, \mathcal{E}_{13}, \qquad
  \mathcal{E}_{3} = \case{1}{\sqrt{3}}\, \mathcal{E}_{12}, }
    \label{eq_app_Weyl_b}
\end{eqnarray}
with similar equations for the frame components of $H_{\alpha\beta}$. It then
follows from \eref{eq_app_Weyl_defn}--\eref{eq_app_Weyl_b} that
\begin{equation}
  \mathcal{W}^{2} = \mathcal{E}_{+}^{2} + \mathcal{E}_{-}^{2}
    + \mathcal{E}_{\times}^{2} + \mathcal{E}_{2}^{2} + \mathcal{E}_{3}^{2}
    + \mathcal{H}_{+}^{2} + \mathcal{H}_{-}^{2} + \mathcal{H}_{\times}^{2}
    + \mathcal{H}_{2}^{2} + \mathcal{H}_{3}^{2} . \label{eq_app_Weyl_c}
\end{equation}
With the frame choice made in appendix~A, equations (1.101) and (1.102) in WE
for $E_{\alpha\beta}$ and $H_{\alpha\beta}$, in conjunction with
\eref{eq_app_eveqn_frame1}--\eref{eq_app_eveqn_frame5}, lead to
\begin{eqnarray}
  \eqalign{
  \mathcal{E}_{+} = \Sigma_{+}(1 + \Sigma_{+}) - \Sigma_{-}^{2}
    - \Sigma_{\times}^{2} + \case{1}{2} \Sigma_{2}^{2}
    + 2( N_{-}^{2} + 3 A^{2} ), \\
  \mathcal{E}_{-} = \Sigma_{-}(1 - 2\Sigma_{+}) + \case{\sqrt{3}}{2}\,
    \Sigma_{2}^{2} + 2\sqrt{3}(N_{-}^{2} - A^{2}), \\
  \mathcal{E}_{\times} = \Sigma_{\times}(1 - 2\Sigma_{+})
    + 8 N_{-} A, \\
  \mathcal{E}_{2} = \Sigma_{2}(1 + \Sigma_{+} + \sqrt{3}\, \Sigma_{-}), \qquad
  \mathcal{E}_{3} = -\sqrt{3}\, \Sigma_{2} \Sigma_{\times},}
    \label{eq_app_Weyl_d}
\end{eqnarray}
and
\begin{eqnarray}
  \eqalign{
  \mathcal{H}_{+} = -3 \Sigma_{-} N_{-} - 3\sqrt{3}\, \Sigma_{\times} A, \\
  \mathcal{H}_{-} = -3 \Sigma_{+} N_{-} - 2\sqrt{3}\, \Sigma_{-} N_{-}
    + \Sigma_{\times} A, \\
  \mathcal{H}_{\times} = -( \Sigma_{-} + 3 \sqrt{3}\, \Sigma_{+})A
    - 2\sqrt{3}\, \Sigma_{\times} N_{-}, \\
  \mathcal{H}_{2} = \sqrt{3}\, \Sigma_{2} N_{-}, \qquad
  \mathcal{H}_{3} = -4 \Sigma_{2} A . } \label{eq_app_Weyl_e}
\end{eqnarray}

%%%%%%%%%%%%%%%%%%%%%%%%%%%%%%%%%%%%%%%%%%%%%%%%%%%%%%%%%%%%%%%%%%%%%%%%%%%

\section*{References}

\begin{harvard}

\item[] Belinskii V A, Khalatnikov I M and Lifshitz E M 1970 Oscillatory
  approach to a singular point in the relativistic cosmology
  \textit{Adv.\ Phys.}\ \textbf{19} 525--73

\item[] Berger B K, Isenberg J and Weaver M 2001 Oscillatory approach to the
  singularity in vacuum spacetimes with $T^{2}$ isometry
  \textit{Phys.\ Rev.\ D} \textbf{64} 084006

\item[] Ellis G F R and MacCallum M A H 1969 A class of homogeneous
  cosmological models \textit{Commun.\ Math.\ Phys.}\ \textbf{12} 108--41

\item[] van Elst H, Uggla C and Wainwright J 2002 Dynamical systems approach
  to $G_{2}$ cosmology \CQG \textbf{19} 51--82

\item[] Hewitt C G 1991 An investigation of the dynamical evolution
  of a class of Bianchi VI$_{-1/9}$ cosmological models \textit{Gen.\ Rel.\
  Grav.}\ \textbf{6} 691--712

\item[] Hewitt C G and Wainwright J 1992 Dynamical systems approach to tilted
  Bianchi cosmologies: irrotational models of type V \textit{Phys. Rev. D}
  \textbf{46} 4242--52

\item[] Hewitt C G and Wainwright J 1993 A dynamical systems approach to
  Bianchi cosmologies: orthogonal models of class B \CQG \textbf{10} 99--124

\item[] Hewitt C G, Wainwright J and Bridson R 2001 The asymptotic regimes
  of tilted Bianchi II cosmologies \textit{Gen.\ Rel.\ Grav.}\
  \textbf{33} 65--94

\item[] Horwood J T, Hancock M J, The D and Wainwright J 2002 Late-time
  asymptotic dynamics of Bianchi VIII cosmologies (submitted to \CQG)

\item[] LeBlanc V G 1997 Asymptotic states of magnetic Bianchi I cosmologies
  \CQG \textbf{14} 2281--301

\item[] Lim W C 2002 private communication

\item[] MacCallum M A H 1973 \textit{Cargese Lectures in Physics} vol~6
  \textit{Lectures at the Int.\ Summer School of Physics (Cargese, 1971)}
  ed E~Schatzmann

\item[] Misner C W 1969 Mixmaster universe \textit{Phys.\ Rev.\ Lett.}\
  \textbf{22} 1071--4

\item[] Wainwright J 1979 A classification scheme for non-rotating
  inhomogeneous cosmologies \JPA \textbf{12} 2015--29

\item[] Wainwright J and Ellis G F R (eds) 1997 \textit{Dynamical Systems in
  Cosmology} (Cambridge: Cambridge University Press)

\item[] Wainwright J, Hancock M J and Uggla C 1999 Asymptotic self-similarity
  breaking at late times in cosmology \CQG \textbf{16} 2577--98

\item[] Weaver M 2002 private communication

\end{harvard}

\end{document}